\documentstyle[preprint,aps,pre,epsfig,amsmath]{revtex}
\begin{document}
\title{Phase Transitions of Soft Disks in External Periodic
Potentials: A Monte Carlo Study}

\author{W. Strepp$^1$, S. Sengupta$^2$, P. Nielaba$^1$}

\address{$^1$ Physics Department, University of Konstanz, 
Fach M 691, 78457 Konstanz, Germany \\
$^2$ S.N. Bose National Centre for Basic Sciences,
Block JD, Sector III, Salt Lake, Calcutta 700098, India}                       
\maketitle

\date{\today}

\begin{abstract}
The nature of freezing and melting transitions for a system of
model colloids interacting by a DLVO potential
in a spatially periodic external potential is studied
using extensive Monte Carlo simulations. Detailed finite size 
scaling analyses of various thermodynamic quantities like the 
order parameter, its cumulants etc. are used to map the phase 
diagram of the system for various values of the reduced screening
length $\kappa a_{s}$ and 
the amplitude of the external potential. We find clear indication
of a reentrant liquid phase over a significant region of the parameter space.
Our simulations therefore show that the system of soft disks behaves in a
fashion similar to charge stabilized colloids which are known to undergo
an initial freezing, followed by a re-melting transition 
as the amplitude of the imposed, modulating 
field produced by crossed laser beams is steadily increased.
Detailed analysis of our data shows several features consistent with a
recent dislocation unbinding theory of laser induced melting.
\end{abstract}


\newpage
\section{Introduction}
The liquid-solid transition in two dimensional systems of particles 
under the influence of
external modulating potentials has recently attracted a fair amount of 
attention from experiments \cite{CAK,CAC,LA,bech,BWL,BBL,maret}, 
theory~\cite{CKS,FNR}
and computer simulations~\cite{CKSS,DK,DSK,DSK2}. This is partly 
due to the fact that well controlled, clean experiments can be performed 
using colloidal particles~\cite{col} confined between glass plates (producing
essentially a two-dimensional system) and subjected to a spatially periodic 
electromagnetic field generated by interfering two, or more, crossed 
laser beams.  One of the more surprising results of these studies, where a  
commensurate, one dimensional, modulating potential is imposed, 
is the fact that there
exist regions in the phase diagram over which one observes 
reentrant~\cite{bech,BWL,BBL} 
freezing/melting behavior. As a function of the laser field intensity the
system first freezes from a modulated liquid to a two dimensional triangular 
solid. A further increase of the intensity confines the particles strongly 
within the troughs of the external potential, suppressing fluctuations
perpendicular to the troughs, which leads to an uncoupling of
neighboring troughs and to re-melting.


Our present understanding of this curious phenomenon has come from early 
mean field density functional~\cite{CKS} and more recent dislocation
unbinding~\cite{FNR} calculations. The mean field theories neglect 
fluctuations and therefore cannot explain reentrant behavior. The order
of the transition is predicted to be first order for small laser field 
intensities, though for certain combinations of external potentials (which 
includes the specific geometry studied in the experiments and in this 
paper) the transition may become second order after going through a 
tricritical point. In general, though mean field theories are applicable 
in any dimension, the results are expected to be accurate only for higher
dimensions and long ranged potentials. The validity of the predictions of 
such theories for the system under consideration is, therefore, in doubt.  

A more recent theory~\cite{FNR} extends the dislocation unbinding mechanism
for two-dimensional melting~\cite{NH,KTHNY} to systems under external 
potentials. For a two-dimensional triangular solid subjected to an external 
one-dimensional modulating potential, the only dislocations involved are 
those which have their Burger's vectors parallel to the troughs of the 
potential. The system, therefore, maps onto an anisotropic, scalar Coulomb 
gas (or XY model)~\cite{FNR} in contrast to a {\em vector} Coulomb 
gas~\cite{NH,KTHNY} for 
the pure $2D$ melting problem. Once bound dislocation pairs are integrated 
out, the melting temperature is obtained as a function of the renormalized 
or ``effective'' elastic constants which depend on external parameters like 
the strength of the potential, temperature and/or density. Though 
explicit calculations are possible only near the two extreme limits of
zero and infinite field intensities, one can argue effectively that a 
reentrant melting transition is expected on general grounds  
quite independent of the detailed nature of the interaction potential 
for any two-dimensional 
system subject to such external potentials. The actual extent of this region
could, of course, vary from system to system. In addition, these authors 
predict that the autocorrelation function of the Fourier components of 
the density (the Debye-Waller correlation function) decays algebraically in 
the solid phase at the transition with an universal exponent which
depends only on the geometry 
and the magnitude of the reciprocal lattice vector. 

Computer simulation results in this field 
have so far been inconclusive. Early simulations~\cite{CKSS} involving 
colloidal particles
interacting via the Derjaguin, Landau, Verwey and Overbeek (DLVO) 
potential~\cite{col} found a large reentrant region
in apparent agreement with later experiments. On closer scrutiny, though,
quantitative agreement between simulation and experiments 
on the same system (but with slightly different parameters) appears to be 
poor~\cite{BBL}. Subsequent 
simulations~\cite{DK,DSK,DSK2} have questioned
the findings of the earlier computation and the calculated phase diagram 
does not show a significant reentrant liquid phase. 

Motivated, in part,
by this controversy, in Ref.~\cite{SSN} we have recently investigated 
the freezing/melting behavior of a two dimensional hard disk system 
in an external potential.  
The pure hard disk system is rather well
studied~\cite{WB,henning,Jas,SNB} by now and the nature of the melting 
transition in the 
absence of external potentials reasonably well explored. Also, there 
exist colloidal systems with hard interactions~\cite{col}
so that, at least in principle, actual experiments using this system 
are possible. Finally, a hard disk simulation
is relatively cheap to implement and one can make detailed
studies of large systems without straining computational resources. 
By these calculations we obtained a
clear signature of a reentrant liquid phase showing that this
phenomenon is indeed a general one as indicated in Ref.~\onlinecite{FNR}.

In the present paper we studied a system of particles interacting by
a DLVO potential in a external periodic potential, motivated on one
hand whether this reentrance scenario is dependent on the range of
interaction, and on the other hand to compare it with experimental results
\cite{BBL}.

The phase diagram has been computed by an application of 
finite size scaling methods similar to the methods used in
our study of the hard disk systems in external potentials~\cite{SSN}.

The rest of our paper is organized as follows. In Section II we specify the 
model and the simulation method including details of the finite size analysis
used. In Section III we present our results for both zero and non-zero
external potential, in particular results for the order parameter and its
cumulants with a discussion on finite size effects. We also present other
quantities like order parameter susceptibility, correlation functions
and heat capacity, which further illustrates the nature of the phase 
transitions in this system.  
In Section IV we discuss our work in relation to the existing literature 
on this subject, summarize and conclude.

\section{Model and method}

\subsection{The Model}
\subsubsection{Potentials}
We study a system of N soft disks in a two dimensional
box of fixed volume
interacting with the DLVO pair potential $\phi(r_{ij})$~\cite{col}
between particles $i$ and $j$ with distance $r_{ij}$,

\begin{equation}
\label{interaction}
\phi(r_{ij})= \frac {(Z^{*}e)^{2}}{4\pi\epsilon_{0}\epsilon_{r}}
\left(\frac{\exp(\kappa R)}{1+\kappa R}\right)^{2}\frac{\exp (-\kappa
  r_{ij})}{r_{ij}} \ \ ,
\end{equation}  
where $R$ is the radius of the particles,
$\kappa=\sqrt{\frac{e^{2}}{\epsilon_{0}\epsilon_{r}k_{B}T}
\sum_{i}n_{i}z_{i}^{2}}$ is the inverse
debye screening length, $Z^{*}$ is the effective surface charge, and
$\epsilon_{r}$ is
the dielectric constant of water. We used $\epsilon_{r}=78$ and,
unless otherwise indicated, $2R=1.07\mu m$ and $Z^{*}=7800$.
Additionally we chose a temperature of $T=293.15 K$, and the particle
density such that the particle spacing of an ideal lattice
is $a_{s}=2.52578\mu m$. We then obtain different values for the
reduced inverse screening length $\kappa a_{s}$ by
varying $\kappa$ as needed.
In our simulation we set $2R$ to be the unit length.
The potential in eq. (\ref{interaction}) mainly depends on the 
value of $\kappa a_{s}$, so all
features found for this system should be valid also for slightly
different values of the other parameters mentioned above.

In addition a particle with coordinates $(x,y)$ is exposed to an
external periodic potential of the form:
\begin{equation}
\label{field}
V(x,y) = V_0\, \sin\, (2 \pi\, x/d_0)
\end{equation}
The constant $d_0$ in Eq.(\ref{field}) is chosen such that, for a 
density $\rho= N/S_xS_y$, the modulation is commensurate to a triangular
lattice of disks with nearest neighbor distance $a_s$: $d_0= a_s \sqrt{3}/2$.

The main parameters which define our system are
$\kappa a_s$ and the reduced potential strength $V_0/k_B T = 
V_0^\ast$, 
where $k_B$ is the Boltzmann constant. 
\subsubsection{box geometry}
All of the data (unless otherwise indicated) presented for 
$V_{0}^{*}<0.2$ are obtained by a simulation in a rectangular box
of size $S_{x}*S_{y}$ ($S_{x}/S_{y}=\sqrt{3}/2$) and periodic boundary
conditions in x- and y-direction, i.e. exactly as in \cite{SSN}.
We will refer to it as 'fixed box geometry' in the rest of the paper.

\begin{figure}[hbtp]
\begin{picture}(0,80)
\put(0,0) {\psfig{figure=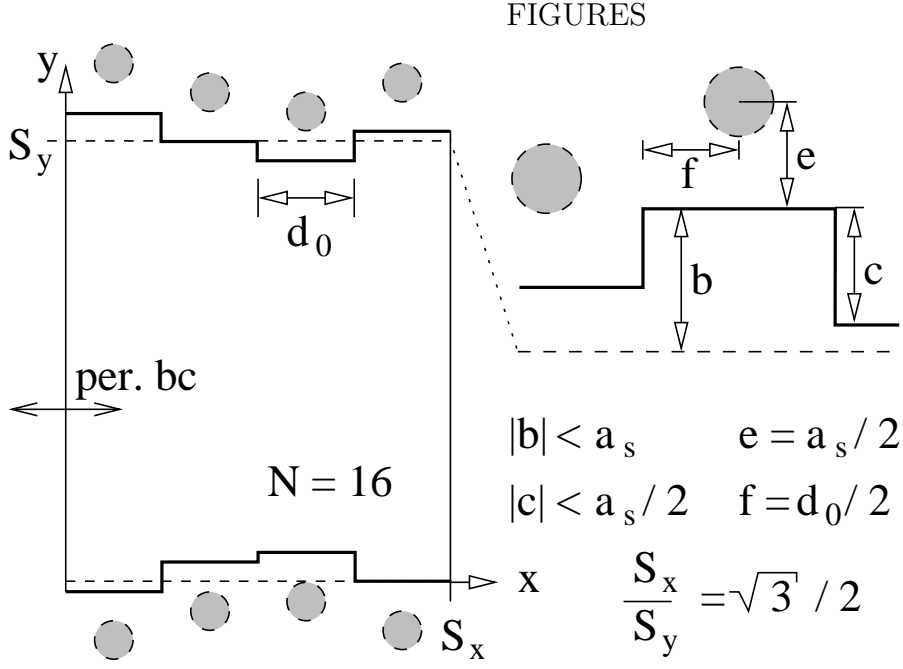,width=120mm}}
\end{picture}
\vskip .5 cm
\caption[]
{
Schematic picture of the simulation box geometry used mainly for
$V_{0}^{*}\ge0.2$.}
\label{geometry}
\end{figure}

For $V_0^* \neq 0$ the external potential modulates the structure
of the fluid and the particles form troughs oriented in the $y$-direction. 
In order to avoid unphysical results, 
for $V_{0}^{*}\ge0.2$ we mainly used a box with periodic boundary
conditions in x-direction and {\em movable walls}
in y-direction, see Fig.~\ref{geometry}, and we will call this 'variable
box geometry'. The simulation box volume is fixed as well as
the side length $S_x$, but
in $x$-direction the box is divided into slabs of width $d_0$, 
centered around the minima of the external periodic potential.
The wall at the end of each slab can move at most $a_{s}$ 
upwards or downwards around its equilibrium position: $|b|< a_{s}$,
such that each slab has variable length between $S_y-2a_s$ and
$S_y+2a_s$ in $y$-direction. The averaged box geometry
still is $S_{x}/S_{y}=\sqrt{3}/2$ as for the fixed one.
The constraint for neighboring walls is to have a distance less than
$a_{s}/2$: $|c|<a_{s}/2$. The walls are hard, so no particle can cross
them. This is indicated as thick solid line in Fig.~\ref{geometry}.
To accommodate the particles in the box as well as possible, additional
'boundary' particles were placed in center ($f=d_{0}/2$)
behind each wall at a distance of $e=a_{s}/2$. The boundary particles
interact with the particles in the box by the usual DLVO potential,
but do not interact with each other. The motion of the walls is done
with a Monte-Carlo procedure and keeps the volume constant, so we are
still in the NVT (but variable shape) ensemble.

The movable walls are chosen to give the system additional degrees of
freedom to relax internal stress.
We were lead to this geometry by some unphysical
results when using the fixed box geometry and higher particle numbers
($N\ge 4096$). For a detailed discussion see end of section \ref{extsec}.

\subsection{The Method}

\subsubsection{Numerical Details}
We perform NVT Monte Carlo (MC) simulations~\cite{metro,LB}
for the system with 
interactions given by Eqs.~(\ref{interaction}) and (\ref{field})
for various values of  $V_0^\ast $ and $\kappa a_s$. 

\begin{figure}[hbtp]
\begin{picture}(0,80)
\put(0,0) {\psfig{figure=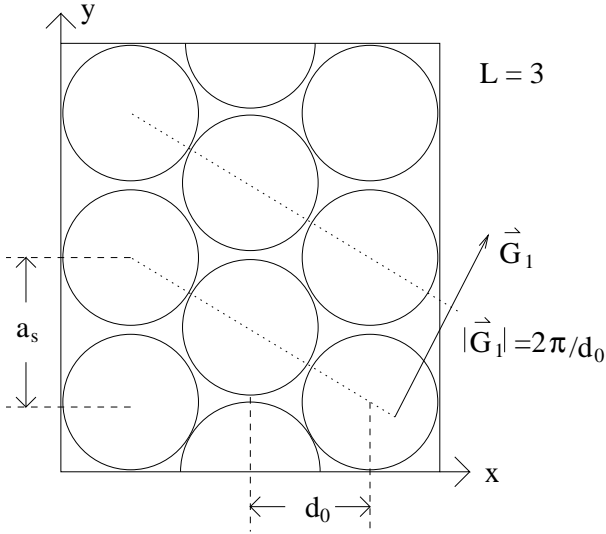,width=80mm}}
\end{picture}
\vskip .5 cm
\caption[]
{
~Schematic picture of the system geometry showing the direction
$\vec{G}_{1}$ along which crystalline order develops at the transition
modulated liquid to solid.
The four vectors obtained by rotating $\vec{G}_1$ anti-clockwise by 
$60^\circ$ and/or reflecting about the origin are equivalent.
}
\label{g1}
\end{figure}

Averages $<\cdot>$ of observables have been obtained with the canonical 
measure. In order to obtain thermodynamic quantities for a range of system
sizes, we analyzed various quantities within subsystems
and used $<\cdot>_L$ to denote averages 
in it. The subsystems are of size $L_x \times L_y$ where $L_x$
and $L_y$ are chosen as $L_y = L a_s$ and $L_x = L_y \sqrt{3}/2=Ld_{0}$
consistent with the geometry of the triangular lattice.
A subbox of size $L=3$ as shown in Fig.~\ref{g1} contains in average
$N_{L}=L^{2}=9$ particles. 

Most of the simulations described below have been done for a total
system size of $N=1024$ 
and $N=4096$ particles, additional ones with $N=400$. Phase transitions
have been studied in most cases by starting in the ordered solid
and increasing $\kappa a_s$ for fixed $V_0^\ast$. Runs where 
$\kappa a_s$ is decreased were also performed for comparison. 

A typical simulation run with $10^7$ Monte Carlo steps (MCS) per particle
(including 3 $\times 10^6$ MCS for relaxation)
took about 50 CPU hours on a PII/500 MHz PC.
In addition to ordinary (local) MC moves
we also used `trough moves', 
by which particle placements in neighboring troughs are tried.
Besides producing faster equilibration, including such moves ensures that 
at high $V_{0}^{*}$ the formation of dislocations 
is not artificially hindered since particles can in effect bypass each other
more easily --- this is very unlikely with purely local MC moves.

\subsubsection{Observables}

\noindent
The potential energy of the system per particle $\varepsilon$ is
computed by:
\begin{equation}
\varepsilon^* = \frac{1}{Nk_BT} 
\sum_{i=1}^N \left[\sum_{j > i} \phi(r_{ij}) +V(x_i,y_i)  \right]
\end{equation}
\noindent
and the heat capacity per particle from the fluctuations of $\varepsilon^{*}$:
\begin{equation}
\frac{c_{V}}{k_{B}}=N<(\varepsilon^{*} - <\varepsilon^{*}>)^{2}>
\end{equation}
\noindent

The nature of the fluid-solid phase transition in two dimensions 
has been a topic of controversy throughout the last forty 
years~\cite{henning,Jas,SNB,aldwain,NH,KTHNY,ecstrand,SNRB}.
It is well known that true long range positional order is absent in 
the infinitely
large system due to low energy long wavelength excitations so that 
translational correlations decay algebraically. According to the 
dislocation unbinding mechanism~\cite{NH,KTHNY,ecstrand} the two dimensional
solid (with quasi long ranged positional and long ranged orientational
order) first melts into a 
``hexatic'' phase with no positional order but with quasi long ranged
orientational order signified by an algebraic decay of
bond-orientational correlation.
A second KT transition, driven by disclination unbinding, leads to 
melting of the hexatic into the liquid, where both the orientational
and positional order is short ranged.
Therefore a useful order parameter in zero external field is the
orientational order parameter. For a particle $j$ located at $\vec{r}_{j}$ 
we define the local orientational order:
$$\psi_{6,j}=\frac{1}{N_{b}}\sum_{l=1}^{N_{b}}
e^{i6\theta_{lj}}$$
where $N_{b}$ is the number of nearest neighbors, and
$\theta_{lj}$ the angle between the axis $\vec{r}_{l}-\vec{r}_{j}$ 
and an arbitrary reference axis.
For the total system we use:
$$
\psi_{6}=\left|\frac{1}{N}\sum_{j=1}^{N}\psi_{6,j}\right|
$$
and as orientational correlation function:
$$
g_{6}(r_{ij})=|<\psi_{6,i}^{*}\psi_{6,j}>|
$$ 
 
In an external periodic field given by Eq.(\ref{field}), however, 
the bond orientational order parameter is nonzero even in the fluid 
phase~\cite{FNR,DSK}.
This is because for  $V_{0}^{\ast} \neq 0$ we have now a ``modulated'' liquid, 
in which local hexagons consisting of the six nearest neighbors of a particle 
are automatically  oriented by the external field. Thus $<\psi_{6}>$
is nonzero both in the (modulated) liquid and the crystalline phase and it
cannot be used to study phase transitions in this system.
The order parameters corresponding to a solid phase are the Fourier 
components of the (non-uniform) density $\rho(\vec{r})$ calculated at the 
reciprocal lattice points $\lbrace \vec{G} \rbrace$. This (infinite) set 
of numbers are all zero (for $\vec{G} \neq 0$ ) in an uniform liquid 
phase and nonzero in a solid. We restrict ourselves to the star consisting 
of the six smallest reciprocal lattice vectors of the two-dimensional 
triangular lattice. In the modulated liquid phase that is relevant to our 
system, the Fourier components corresponding to two out of these six vectors,
viz. those in the direction perpendicular to the troughs of the external 
potential, are nonzero~\cite{CKS}. The other four components of this set
consisting of the one in the direction $\vec{G}_1$ (as defined for the ideal
crystal in Fig.~\ref{g1}), and those 
equivalent to it by symmetry, are zero in the (modulated) liquid and 
nonzero in the solid (if there is true long range order).
We therefore use the following order parameter: 
$$\psi_{G_1} = \left|\frac{1}{N}\sum_{j=1}^N \exp (i \vec{G}_1 \cdot \vec{r}_j)\right|$$
where $\vec{r}_j$ is the position vector of the $j^{th}$ particle.
The corresponding susceptibility $\chi_{G_1}$ is:
\begin{equation}
\label{suszep}
k_BT \chi_{G_1} = L^2 \left[ 
\left<\left(\psi_{G_1}\right)^2\right> - 
\left<\psi_{G_1}\right>^2 
\right]
\end{equation}
To measure the positional correlation, we chose the Debye-Waller
correlation function which we define as follows:
$$C_{\vec{G}_{1}}(\vec{R})=|<e^{i\vec{G}_{1}(\vec{u}(\vec{R})-\vec{u}(0))}>|$$
\noindent
where $\vec{R}$ points to the elementary cell of the ideal
lattice, and $\vec{u}(\vec{R})$ is the deviation of the actual particle
position from the ideal lattice:
$\vec{r}=\vec{R}+\vec{u}(\vec{R})$.      
In this case we have chosen the direction of $\vec{R}$ to lie along the
$y$ axis (i.e. along the troughs of the potential). In the solid we
expect this quantity do decay algebraically,
i.e. $C_{\vec{G}}(y)\propto1/y^{\eta_{\vec{G}}}$ \cite{FNR,NH}, where
$\eta_{\vec{G}}$ depends on the elastic constants.

$\psi_{G_{1}}$ is sensitive to the phase transition where positional order
is lost. Therefore, when decreasing $V_{0}^{*}$ we expect 
the phase boundary to converge to the corresponding transition
at zero external potential. But in contrast to $V_{0}^{*}\not=0$ where 
the crystal is oriented by the external potential, at $V_{0}^{*}=0$
it is only weakly fixed by the boundary conditions and can
start to rotate, so we can not apply $\psi_{G_{1}}$ there.
For this purpose we use a slightly modified positional order parameter 
$\tilde{\psi}_{G_{1}}$ at $V_{0}^{*}=0$: the phase information of 
$\psi_{6}$ (of course, before taking the absolute value) is used to determine
the orientation of the crystal, and then a tilted coordinate system is
used to compute $\tilde{\psi}_{G_{1}}$. We applied the same method
when calculating $C_{\vec{G}_{1}}(y)$ at $V_{0}^{*}=0$. 

We have determined phase transition points by the
order parameter cumulant intersection method.
The fourth order cumulant $U_L$ of the order parameter distribution 
is given by \cite{KB}:
\begin{equation}
U_L(V_0^\ast,\kappa a_{s}) = 1 - \frac{<\psi_{x}^4>_L}{3~<\psi_{x}^2>_L^2}
\end{equation}
In order to distuingish between the cumulants of $\psi_{6}$ and
$\psi_{G_{1}}$, we denote them with $U_{L,6}$ and $U_{L,G}$, respectively.
In the liquid (short ranged order) $U_{L}\rightarrow 1/3$ and in the
solid (long range order) $U_{L}\rightarrow 2/3$ for
$L\rightarrow\infty$.
In case of a continuous transition close to the transition point
the cumulant is only a function of the ratio of the system size 
$\approx L a_s$ and the correlation length $\xi$: $U_L(L a_s/\xi)$.
Since $\xi$ diverges at the critical point the cumulants for 
different system sizes intersect in one point: 
$U_{L_1}(0)= U_{L_2}(0) = U^\ast$. $U^{*}$ is a non-trivial value,
i.e. $U^{*}\not=1/3$ and $U^{*}\not=2/3$. 
Even for first order transitions these cumulants intersect~\cite{VRSB} though 
the value $U^\ast$ of $U_L$ at the intersection is not
universal any more. The intersection point can, therefore, be taken as 
the phase boundary regardless of the order of the transition.
This is useful since the order of the melting transition in $2D$ 
either in the absence~\cite{henning,Jas,SNB,aldwain,NH,KTHNY,ecstrand,SNRB}
or with~\cite{CKS,FNR,CKSS,DK,DSK,DSK2} external potentials is not 
unequivocally settled. And there is also another aspect in our 
special case: since the positional order correlation is predicted to
decay algebraically in the solid phase (quasi long ranged order),
the whole solid can be seen
as consisting of a line or area of critical points with temperature-
dependent critical indices $\eta_{\vec{G}}(T)$ \cite{NH}. In that case
we expect the cumulants to merge at a nontrivial value at the 
onset of the solid phase
instead of intersecting, yielding a line of intersections.
We indeed observed this behavior for hard disks in high
external potentials \cite{SSN}. For the same reasons the same behavior 
is expected for the $\psi_{6}$-cumulants at the liquid-hexatic
transition and in the hexatic phase\cite{henning}. Also the very
similar 2d-XY-spin model shows this behavior when using the magnetization
as order parameter \cite{DL}. 

Note that though the order parameter  
$<\psi_{G_1}>$ decays to zero with increasing system size 
even in the 2-d solid (assuming quasi long ranged order there),
the cumulants will stay at the non-trivial value regardless of $L$.
So for $L\rightarrow\infty$ there should be a jump
from $1/3$ to this nontrivial value when crossing
the phase boundary from liquid to solid, which underlines
the usefulness of cumulants.

In order to map the phase diagram we systematically vary the
system parameters $V_0^\ast$ and $\kappa a_s$
to detect order parameter cumulant intersection- or merging points which are then 
identified with the phase boundary.

\section{Results and Discussion}

\subsection{Zero external potential}

\noindent
In this section we analyse the system properties for zero 
external field. In particular we present results for the order parameter,
the cumulants, the correlation functions and the heat capacity
for different values of $\kappa a_s$. In these studies we used the
fixed box geometry.

\begin{figure}[hbtp]
\begin{picture}(0,80)
\put(0,0) {\psfig{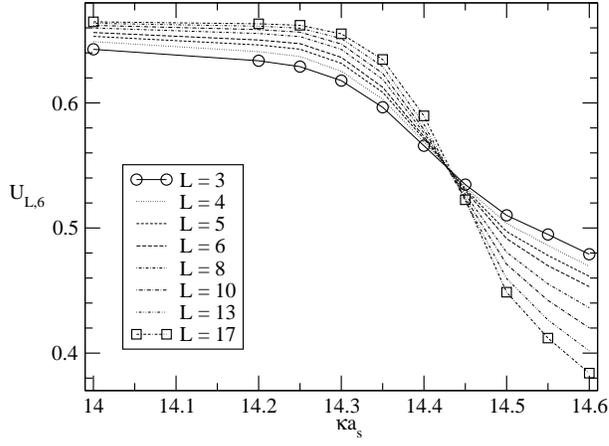}}
\end{picture}
\vskip .5 cm
\caption[]
{
Cumulant of the $\psi_6$ order parameter versus $\kappa a_s$
for various values of the system size $L$
(N=4096, $V_0^{*}$=0).
}
\label{psi60}
\end{figure}

\begin{figure}[hbtp]
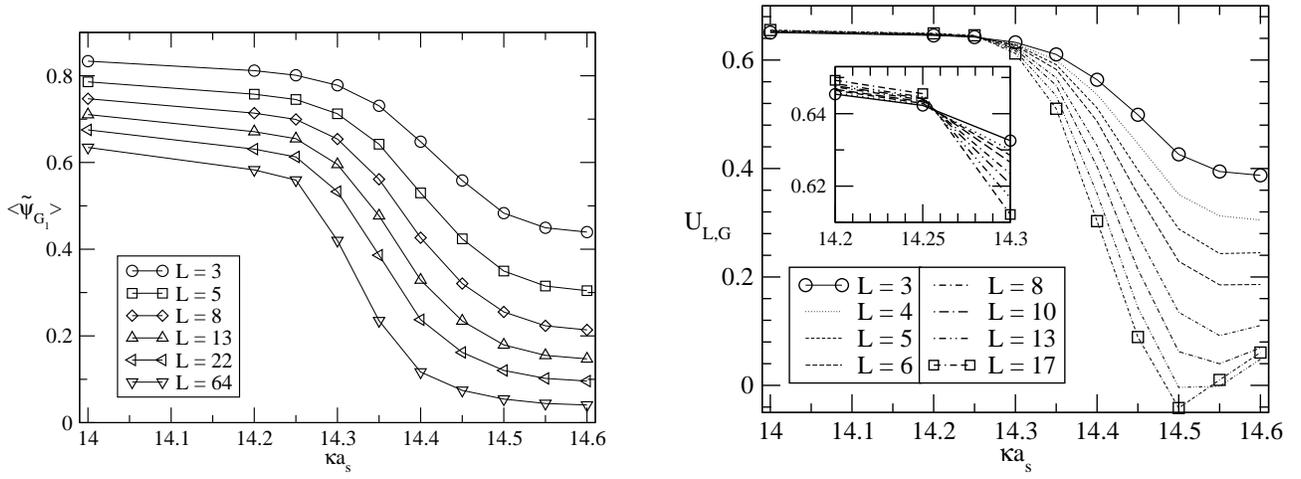

\begin{picture}(0,80)
\put(-10,0) {\psfig{figure=figures/psiGtilde_mittel_V0_N4096_DLVO.eps,width=80mm}}
\put(80,0) {\psfig{figure=figures/psiGtilde_kumul_V0_N4096_DLVO.eps,width=80mm}}
\end{picture}
\vskip .5 cm
\caption[]
{
Average (left) and 
cumulant (right) of the $\tilde{\psi}_{G_1}$ order parameter 
versus $\kappa a_s$ for various values of the system size $L$
(N=4096, $V_0^{*}$=0).
}
\label{psitg0}
\end{figure}

In Fig.~\ref{psi60} the cumulant of the
$\psi_6$ order parameter versus $\kappa a_s$ is shown for different
subsystem sizes. We identify the phase transition value of $\kappa_6 a_s$ 
at about 14.42 by locating the cumulant intersection point.
Since the positional order is not well defined in two-dimensional
systems, the positional order parameter $\tilde{\psi}_{G_1}$ shows a strong
system-size dependency, see Fig.~\ref{psitg0}. The cumulants of 
$\tilde{\psi}_{G_1}$ intersect at a value of 
$\kappa_G a_s \approx 14.25$, which is slightly smaller than $\kappa_6 a_s$.
This is in agreement with a KTHNY two stage melting scenario, in which 
the solid and the fluid phase are separated by a ``hexatic'' region
in the phase diagram, in which the positional order is short ranged
and the bond-orientational order is long ranged. Both effects are
detected by the two order parameters, $\psi_6$ being sensitive for
the bond-orientational order and $\tilde{\psi}_{G_1}$ on the
positional order.
Surprisingly, however, though in
the case of the hexatic phase one expects the $\psi_6$-cumulants
to coincide, they obviously don't in Fig.~\ref{psi60}
(see also \cite{henning}).

\begin{figure}[hbtp]
\begin{picture}(0,80)
\put(0,0) {\psfig{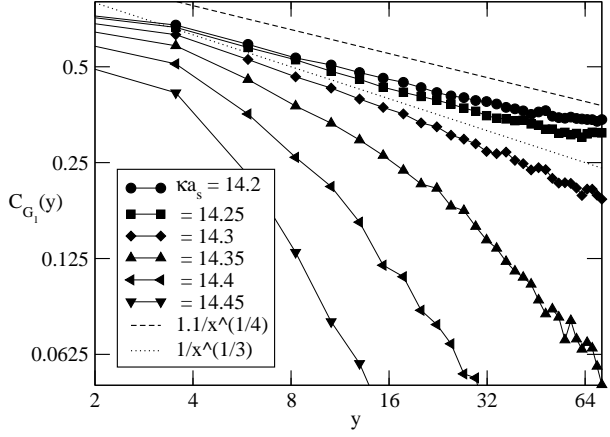}}
\end{picture}
\vskip .5 cm
\caption[]
{
Debye Waller correlation function versus $y$ for various values of
$\kappa a_s$ (N=4096, $V_0^{*}=0$). Dashed line: 
Schematic picture of the functional decay with exponent 1/4, 
dotted line: 
schematic picture of the functional decay with exponent 1/3. 
}
\label{debwal}
\end{figure}
\begin{figure}[hbtp]
\begin{picture}(0,80)
\put(0,0) {\psfig{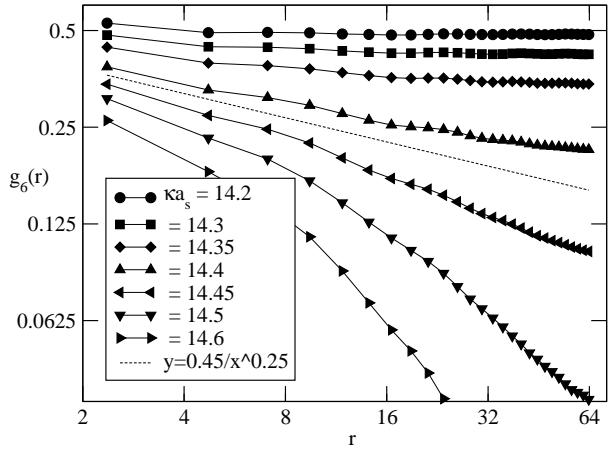}}
\end{picture}
\vskip .5 cm
\caption[]
{
Orientational correlation function versus distance for various values of
$\kappa a_s$ (N=4096, $V_0^{*}=0$). Dashed line: 
schematic picture of the functional decay with exponent 1/4. 
}
\label{orientco}
\end{figure}

The Debye-Waller correlation functions $C_{G_1}(y)$ for
different values of $\kappa a_s$ are shown in Fig.~\ref{debwal}.
At the transition value $\kappa_G a_s=14.25$ for $N=4096$ we find
a power law dependency of $C_{G_1}(y)$ from $y$ with an exponent 
$\eta_{G_{1}}\approx0.28$ which is well within the predicted range
of $[1/4, 1/3]$.
In Fig.~\ref{orientco} the orientational correlation function
versus distance is shown. This function reveals a power law dependency 
of the bond-orientation correlations at $\kappa_6 a_s$, the exponent
value is about 1/4. The value of the exponent $\eta_{6}$ at the transition
has been predicted by the KTHNY theory to be $1/4$~\cite{FNR}, which 
is in agreement with our results.

\begin{figure}[hbtp]
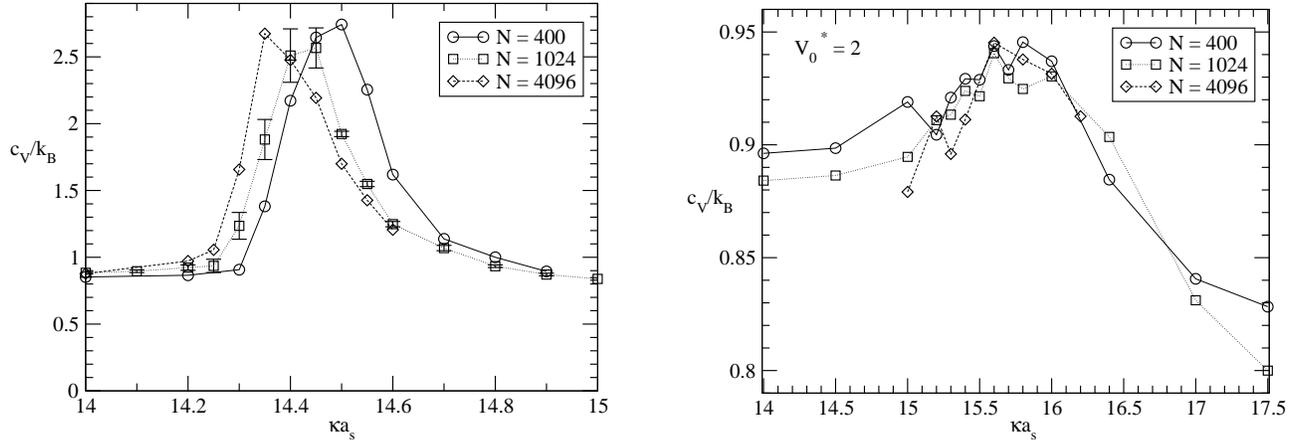

\begin{picture}(0,80)
\put(-10,0) {\psfig{figure=figures/cV_V0_DLVO_vergleich.eps,width=80mm}}
\put(80,0) {\psfig{figure=figures/cV_V2_DLVO2_vergleich.eps,width=80mm}}
\end{picture}
\vskip .5 cm
\caption[]
{
Heat capacity versus $\kappa a_s$ for various numbers of particles.
Left figure: $V_0^*=0$ computed with fixed box geometry, 
right figure: $V_0^*=2$, computed with variable box geometry.
}
\label{cv}
\end{figure}

\begin{figure}[hbtp]
\begin{picture}(0,200)
\put(20,0) {\psfig{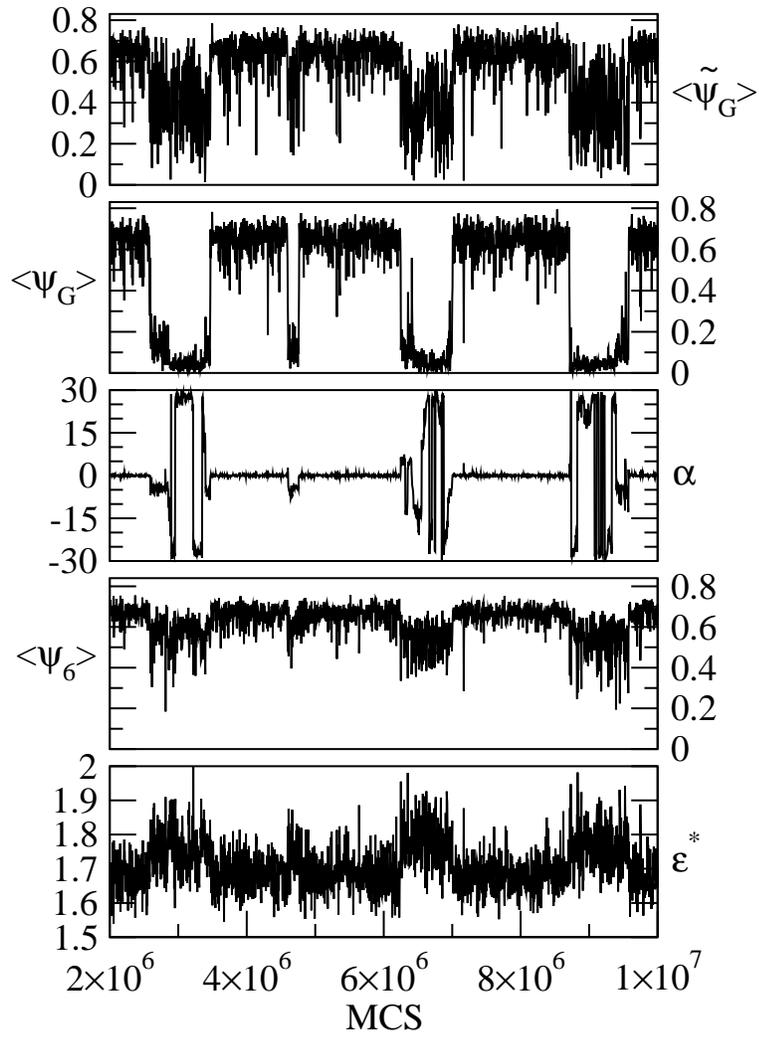}}
\end{picture}
\vskip .5 cm
\caption[]
{
System evolution versus Monte Carlo steps (N=400, $V_0^{*}$=0, $\kappa a_s$=14.4).
From bottom to top: energy $\epsilon^{*}$, $\psi_6$ order parameter, angle $\alpha$
of lattice direction, $\psi_G$ order parameter, $\tilde{\psi}_G$. 
}
\label{sysevol}
\end{figure}

\begin{figure}[hbtp]
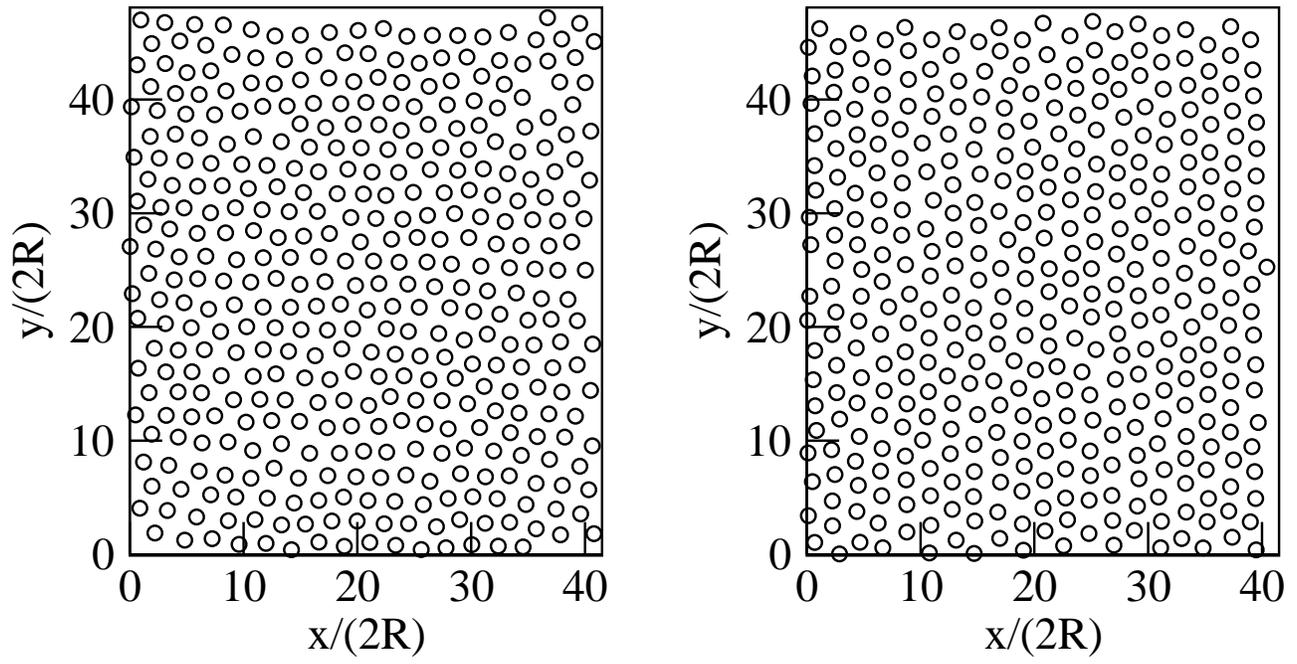

\begin{picture}(0,100)
\put(-10,0) {\psfig{figure=figures/N400_K14.4_V0_DLVO_MCS3000000.eps,width=80mm}}
\put(80,0) {\psfig{figure=figures/N400_K14.4_V0_DLVO_MCS7700000.eps,width=80mm}}
\end{picture}
\vskip .5 cm
\caption[]
{
Configurations after 3.000.000 MCS (left) and 7.700.000 MCS (right)
from the system of Fig.~\ref{sysevol}.
}
\label{configs}
\end{figure}

In Fig.~\ref{cv} (left) we present 
the heat capacity data versus $\kappa a_s$
for different system sizes. It is obvious that the heat capacity
does not show a singularity as would be expected in case of a
first order or a conventional second-order transition. The peak maxima are not very sharp,
but are roughly located
close to the value of $\kappa_6 a_s$, where the $\psi_6$-order parameter 
cumulants intersect. The peak maxima thus do not agree with the cumulant
intersection point of the $\tilde{\psi}_{G_1}$-order parameter, which is again
in agreement with the KTHNY scenario. We note that the identification
of the phase transition point by the heat capacity maxima may result
in misleading results on the location of the transition points in the
phase diagram. 
In particular, for a smaller system of N = 400 particles we find
that configurations wherein the crystal is rotated by a tilt angle
of $\alpha = \pm 30^\circ$ ($\alpha$ extracted from the phase information of
$\psi_{6}$) may be present. Since this is incompatible
with the box geometry it leads to a higher energy and a lower measured
$\psi_{G_1}$. The value of $\tilde{\psi}_{G_1}$, on the other hand,
is not appreciably altered. This is shown in the time evolution of
that system in Fig.~\ref{sysevol}. 
We also show the configuration of 
a tilted crystal with $\alpha=30^{\circ}$ (3.000.000 MCS) in
Fig.~\ref{configs} (left), and of an 'correctly' aligned crystal 
($\alpha=0^{\circ}$, 7.700.000 MCS) (right).

\subsection{External periodic potential}
\label{extsec}
In this section we analyse the system properties in the presence
of a periodic external potential. The studies in this section are
mainly done with 
variable box geometry. Comparative studies with fixed box geometry
show that the new method does not lead to artificial features,
but rather gives improved results. For more details see the
discussion of the Debye-Waller correlation function at the end
of this section.

\begin{figure}[hbtp]
\begin{picture}(0,80)
\put(-10,0) {\psfig{figure=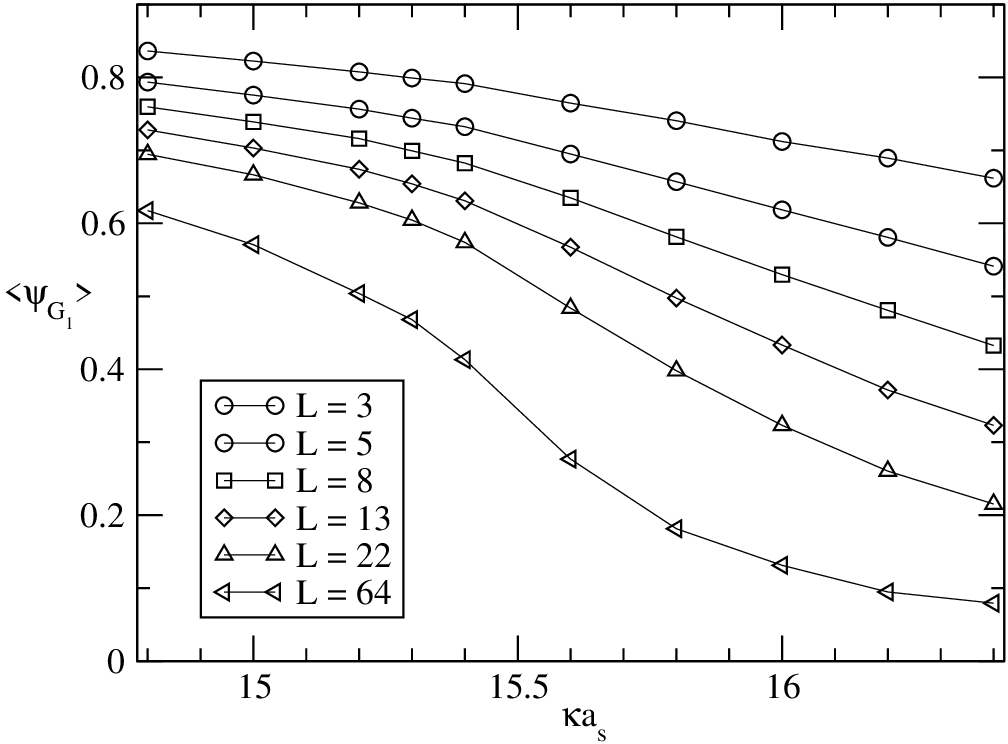,width=80mm}}
\put(80,0) {\psfig{figure=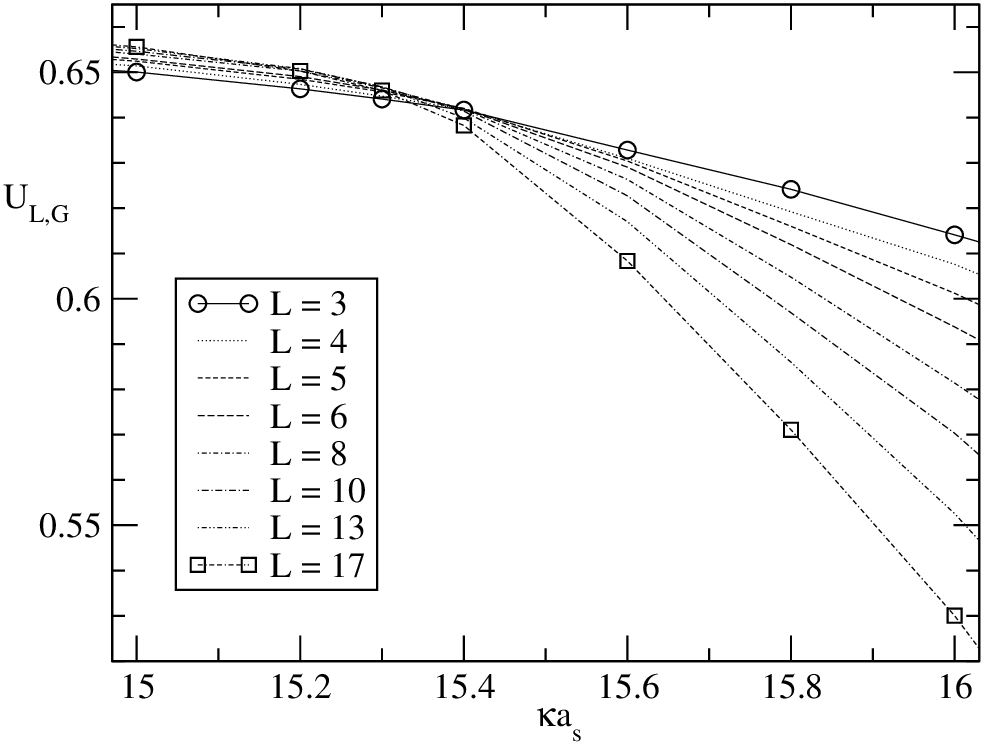,width=80mm}}
\end{picture}
\vskip .5 cm
\caption[]
{
Average (left) and cumulant (right) of the $\psi_{G_1}$ order parameter versus
$\kappa a_s$ for $V_0^* = 2$ and various system sizes $L$ (N=4096).
}
\label{opkumv2}
\end{figure}

\begin{figure}[hbtp]
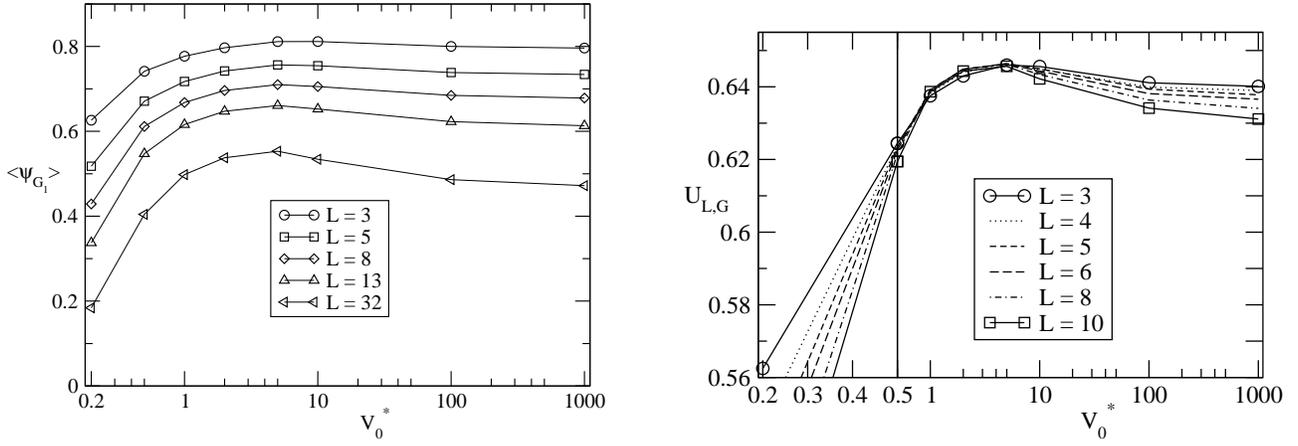

\begin{picture}(0,80)
\put(-10,0) {\psfig{figure=figures/psiG_mittel_kas15.3_N1024_DLVO2.eps,width=80mm}}
\put(80,0) {\psfig{figure=figures/psiG_kumul_kas15.3_N1024_DLVO2.eps,width=80mm}}
\end{picture}
\vskip .5 cm
\caption[]
{
Average (left) and cumulant (right) of the $\psi_{G_1}$-order parameter versus
$V_0^*$ for $\kappa a_s=15.3$ and  various values of L 
and variable box geometry (N=1024).
}
\label{ocka}
\end{figure}

Two examples for the $\kappa a_s$-dependency of the $\psi_{G_1}$-order
parameter and the cumulants for an external potential amplitude
of $V_0^*=2$ and $V_0^*=1000$ are shown in Figs.~\ref{opkumv2} and 
\ref{opkumv1000}. We note that the cumulant intersection, which can be 
clearly identified for $V_0^*=2$ in Fig.~\ref{opkumv2} is developing
towards an intersection ``line'' for $V_0^*=1000$, a behavior which was found in case
of the hard disk system in external potentials~\cite{SSN} as well as in related
systems with a KT transition like the XY-spin-model \cite{DL}.
Another example is shown in Fig.~\ref{ocka}. There $\kappa a_{s}$ is
kept fixed at $15.3$ and $V_{0}^{*}$ is varied. The starting point at
$V_{0}^{*}=0.2$ is in the mod. liquid phase, crosses slightly the
solid ('laser induced freezing'), and re-enters the mod. liquid at
higher $V_{0}^{*}$ ('reentrance'). This is already a first sign of a 
'reentrant' phase transition scenario.
For the phase diagram these cumulant intersection- or merging values
were used.

\begin{figure}[hbtp]
\begin{picture}(0,80)
\put(0,0) {\psfig{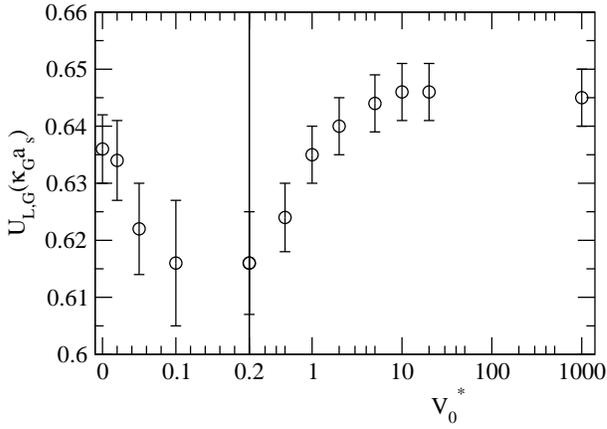}}
\end{picture}
\vskip .5 cm
\caption[]
{
Cumulant intersection values of the $\psi_{G_1}$-order parameter 
versus $V_0^*$ for fixed box geometry (N=1024). The data for variable
box geometry is similar. 
}
\label{cuka}
\end{figure}

In Fig.~\ref{cuka} the cumulant intersection values are shown as a function
of $V_0^\ast$ for fixed box geometry. We observe that $U^\ast$ is
not an universal number but, nevertheless, goes to a limiting value 
for large $V_0^\ast$~\cite{DL}. 

\begin{figure}[hbtp]
\begin{picture}(0,80)
\put(0,0) {\psfig{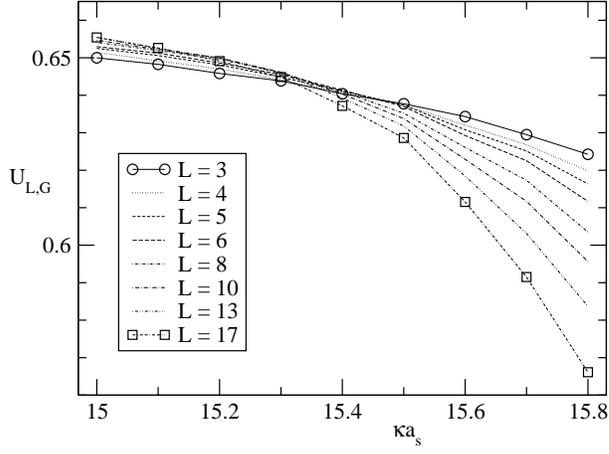}}
\end{picture}
\vskip .5 cm
\caption[]
{
Cumulant of the $\psi_{G_1}$ order parameter versus
$\kappa a_s$ for $V_0^* = 2$ and various system sizes $L$ 
for a density quench path (N=4096).
}
\label{opkum2r}
\end{figure}

The amount of hysteresis effects on the location of the transition
point has been analyzed for the case of $V_0^*=2$ by a time consuming
reverse density quench simulation, in which a path in phase space
was chosen in the direction opposite to the standard path. 
The results of this study are shown in Fig.~\ref{opkum2r}.
Comparing these results with the ones of Fig.~\ref{opkumv2} reveals
quite close agreements showing
that only small hysterese effects are present in the system.

\begin{figure}[hbtp]
\begin{picture}(0,80)
\put(0,0) {\psfig{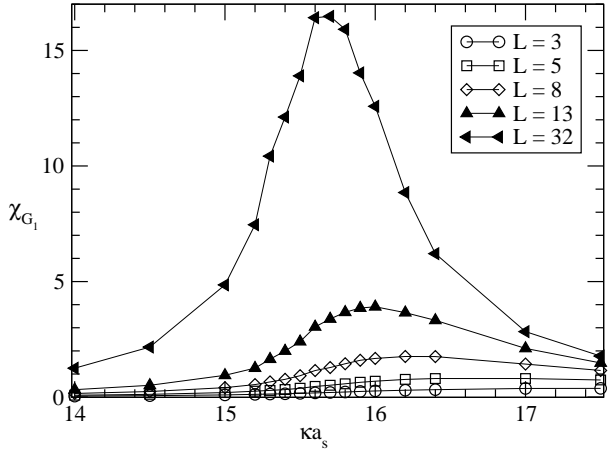}}
\end{picture}
\vskip .5 cm
\caption[]
{
Susceptibility of the $\psi_{G_1}$-order parameter versus
$\kappa a_s$ for $V_0^*$=2 and various values of the system size $L$
(N=1024).
}
\label{sus2}
\end{figure}
\begin{figure}[hbtp]
\begin{picture}(0,80)
\put(0,0) {\psfig{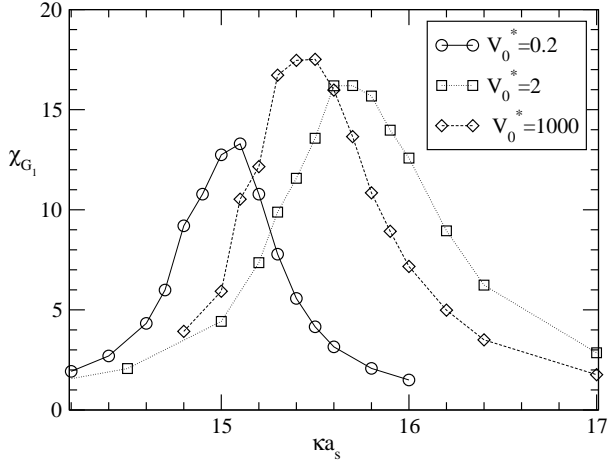}}
\end{picture}
\vskip .5 cm
\caption[]
{
Susceptibility of the $\psi_{G_1}$-order parameter versus
$\kappa a_s$ for various values of $V_0^*$ (L=32, N=1024).
}
\label{susa}
\end{figure}

The $\psi_{G_1}$ order parameter susceptibilities $\chi_{G_1}$
are shown in Fig.~\ref{sus2} versus $\kappa a_s$ for different system sizes. We note
that close to the transition a maximum develops, the value of the maximum
increasing with the system size. 
In Fig.~\ref{susa} the susceptibilities of the largest subsystems 
($L=32$) as functions of $\kappa a_s$ are compared for different values of
$V_0^*$. Clearly the peak position for $V_0^*=2$ is shifted to 
larger values compared to the cases with  $V_0^*=0.2$ and $V_0^*=1000$.
This feature is another sign of a ``reentrant'' phase transition
scenario in the phase diagram. 
Compared to the cumulant intersection values, $\chi_{G_1}$ maxima are located at
slightly higher $\kappa a_{s}$. This may be due to  
finite size effects, which often show the feature that 
phase transition points in finite systems
are shifted to slightly different values depending on the observable
under investigation. In particular one expects (and we get) a shift towards 
parameter values in the disordered region (here a liquid, i.e. higher
$\kappa a_{s}$) for the average order parameter and the susceptibility.

In Fig.~\ref{cv} (right) the heat capacity for
$V_{0}^{*}=2$ and different
system sizes is shown. The peak is nearly independent from system size
and shifted towards the liquid phase with respect to the order
parameter cumulant intersection value, i.e. the same behavior as
for $V_{0}^{*}=0$. 

\begin{figure}[hbtp]
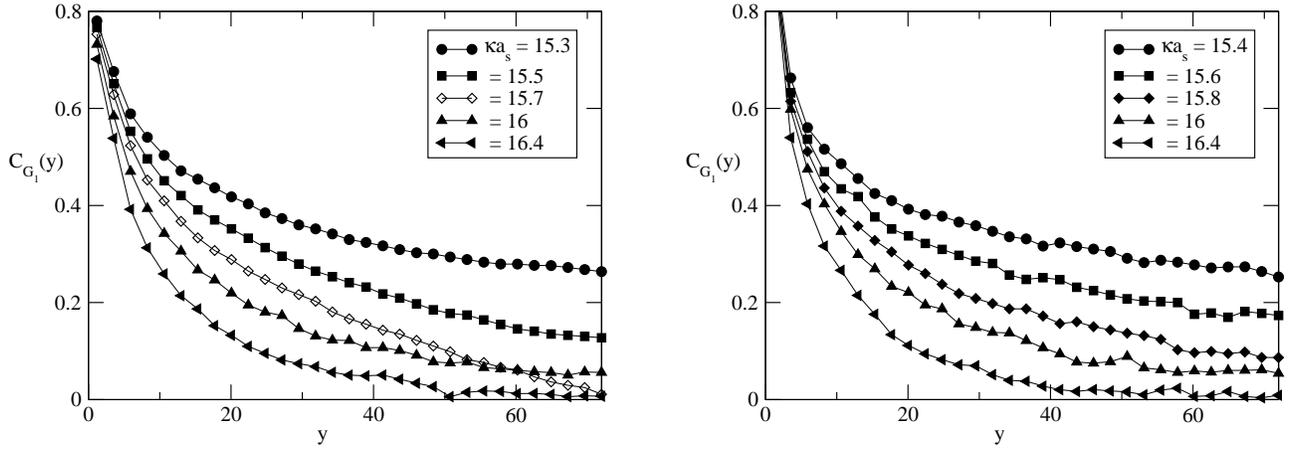

\begin{picture}(0,80)
\put(-10,0) {\psfig{figure=figures/debyeWaller_V2_N4096_DLVO.eps,width=80mm}}
\put(80,0) {\psfig{figure=figures/debyeWaller_V2_N4096_DLVO2.eps,width=80mm}}
\end{picture}
\vskip .5 cm
\caption[]
{
Debye-Waller correlation function $C_{G_1}(y)$ versus $y$ for 
$V_0^*=2$ and various values of $\kappa a_s$ (N=4096).
Correlations for computations with fixed box geometry (left) and
variable box geometry (right).
}
\label{debwal2}
\end{figure}

The advantage of the variable box geometry, especially for large
systems, can be seen best by looking at the Debye-Waller correlation function. 
In Fig.~\ref{debwal2} (left) an example is shown for 
fixed box geometry. The crossing
from the solid phase with an algebraic decay to the mod. liquid with
exponential decay is not monotonic, but at $\kappa a_{s}=15.7$,
$C_{G_{1}}(y)$ drops to zero at $y=S_{y}/2$, which is not physically
meaningful. At a higher value
$\kappa a_{s}=16$ it rises, and then falls again at $\kappa
a_{s}=16.4$, showing a exponential decay as expected. In variable
box geometry this feature doesn't show up, see
Fig.~\ref{debwal2} (right). Here we have a smooth transition from solid- 
to liquid-like behavior. We explain
this strange behavior at $\kappa a_{s}=15.7$ in fixed geometry as follows:
consider a system with $N=10000$ particles. Without dislocations there
will be an ideal lattice with $N_{t}=100$ particles in each of the 
$100$ troughs. Assuming
dislocation unbinding as melting mechanism, consider the existence
of some dislocations in the system with opposite burgers vectors
$\vec{b}=\pm a_{s}\vec{e}_{y}$. One of these dislocations increases
the number of particles in the troughs by one, while the other
decreases it by
one. We can have for example a situation where half of the troughs has $100$
particles, and the other half has either $101$ or $99$ particles. If we now
for simplicity assume that the distance between particles in a row is
$a=L_{y}/N_{t}$, calculating the pair correlation function $g(y)$
along a trough will yield two peaks around $y=L_{y}/2$: one centered exactly at
$y=L_{y}/2$ from the troughs with $N_{t}=100$ ($a=a_{s}$), and the
other centered at
$y=L_{y}/2+a_{s}/2$ due to the troughs with $N_{t}=99$ or
$N_{t}=101$ ($a\not= a_{s}$).
As consequence, $C_{G_{1}}(y=L_{y}/2)$ will be zero.
The same situation in the movable-walls geometry will not show these problems:
the troughs with $101$ particles can expand a bit, while the ones with
$99$ particles can contract. Now in every trough is $a=a_{s}$, and 
$C_{G_{1}}(y=L_{y}/2)$ is not necessarily zero. Also, the formation
of a dislocation pair costs less energy and is closer to the true
infinite system value. The discussion above is 
in some sense similar to the 2d-XY-spin model with
a vortex in the center: in an infinite sample, the spins to the left
and to the right have opposite spin directions, but periodic boundary 
conditions in a finite system
will try to align them, so that the formation of the 
vortex is disturbed. Free boundary conditions won't cause this problem.

In zero external potential with fixed geometry the formation of 
a dislocation pair is not so problematic, since the particles are 
not forced into troughs and
have more degrees of freedom in movement. In the above example one end of a line
of $N_{t}=101$ particles could make a slight shift in $x$-direction
to access more space in $y$-direction.

However, at the first data point in the solid closest to the transition we 
find an algebraic decay $C_{G_{1}}(y)\propto1/y^{\eta_{G_{1}}}$ 
with $\eta_{G_{1}}$ in the range of $0.25\dots0.34$ for
$V_{0}^{*}=0\dots 1000$ and $N=1024$ particles.
In \cite{FNR} $\eta_{G_{1}}$ is predicted to be $1/4$ at the transition.

\subsection{The Phase Diagram}
\label{phdsec}
\begin{figure}[hbtp]
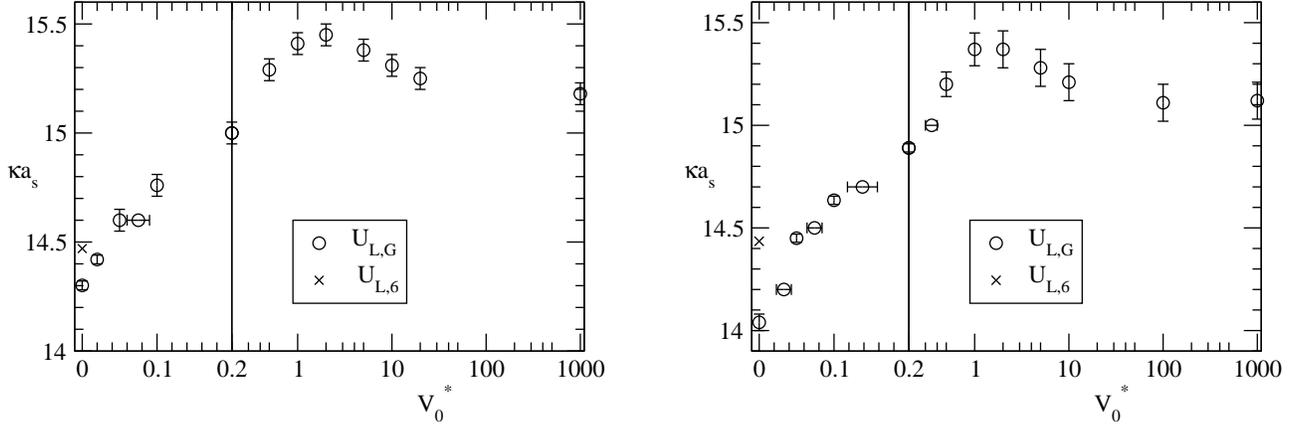

\begin{picture}(0,80)
\put(-10,0) {\psfig{figure=figures/N1024_DLVO_phasediagram.eps,width=80mm}}
\put(80,0) {\psfig{figure=figures/N1024_DLVO2_phasediagram.eps,width=80mm}}
\end{picture}
\vskip .5 cm
\caption[]
{
Phase diagram. The points show the parameters for the cumulant intersection
of the $\psi_{G_1}$-order parameter (N=1024). Left picture: Computations with
fixed box geometry for all $V_0^{*}$. Right picture: computations with
variable box geometry for all $V_0^{*}$.
At $V_0^*=0$ the parameters for the cumulant intersection of the
$\psi_6$ order parameter are shown for comparison.
}
\label{phd}
\end{figure}

For each $\kappa a_{s}$ and $V_0^\ast$ value we computed
cumulants $U_{L,G}$ for a range of subsystem sizes $L$ and located intersection- 
or merging points which we identify with the phase boundary.
We have obtained a detailed phase diagram for $N=1024$ particles 
which is shown in Fig.~\ref{phd} for fixed
box geometry (left) and for variable box geometry (right). We want to
emphasize that there are only slight differences and the general
shape of the phase diagram is the same for both box geometries.
At $V_{0}^{*}=0$ also the $\psi_{6}$ cumulant intersection value
is plotted for comparison.
The values of $\kappa a_{s}$ at the transition initially rise 
and subsequently drop as $V_0^\ast$ increases.
The maximum $\kappa_{G} a_{s}$ values are found for $V_0^\ast \approx 1-2$.
These transition points separate a high density solid
from a low density modulated liquid. Thus, at a properly chosen
$\kappa a_{s}$, we observe an initial freezing transition followed
by a reentrant melting at a higher $V_0^\ast$ value. Such an effect 
had been found earlier in experiments on colloidal systems
in an external laser field~\cite{bech,BWL,BBL}.

\begin{figure}[hbtp]
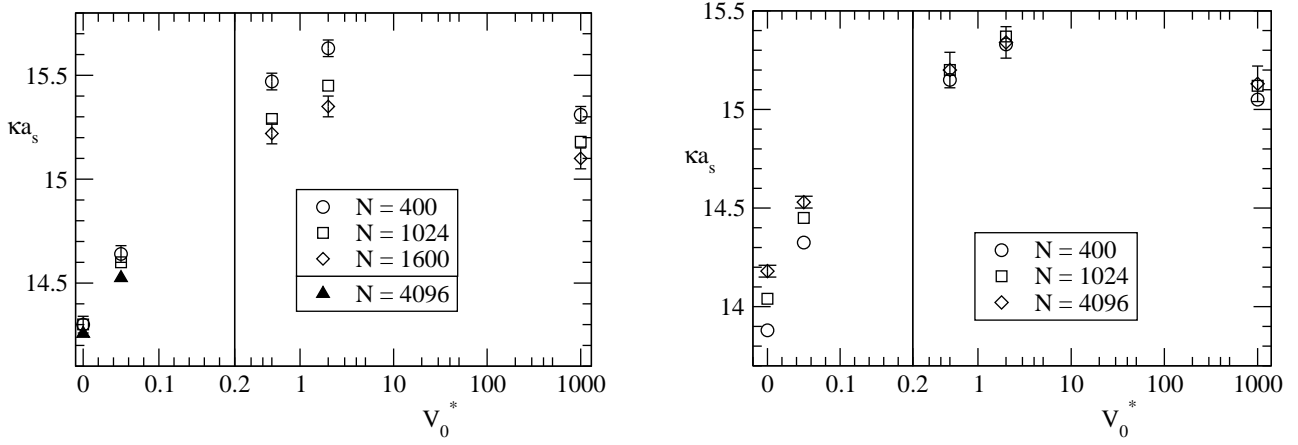

\begin{picture}(0,80)
\put(-10,0) {\psfig{figure=figures/DLVO_phasediagram_vergleich.eps,width=80mm}}
\put(80,0) {\psfig{figure=figures/DLVO2_phasediagram_vergleich.eps,width=80mm}}
\end{picture}
\vskip .5 cm
\caption[]
{
Finite size effects on the phase diagram. 
The points show the parameters for the cumulant intersection
of the $\psi_{G_1}$-order parameter for different total system sizes. 
Left picture: Computations with
fixed box geometry for all $V_0^{*}$, right picture: computations with
variable box geometry for all $V_0^{*}$.
}
\label{phdfs}
\end{figure}

In order to quantify residual finite size effects on the phase diagram,
we have computed the transition points for different total system sizes.
The resulting phase diagrams are shown in Fig.~\ref{phdfs}, again
for fixed (left) and variable box geometry (right). We note that with
increasing system size in fixed box geometry all transition points
are slightly shifted to the solid region, whereas for the variable box
this shift is towards the liquid region. One can see that the shift
for the fixed box is much smaller at low $V_{0}^{*}$, and for the
variable box it is much smaller at medium and high $V_{0}^{*}$.
We also found that the cumulant intersection point smears out
strongly if using the fixed box, $N\ge4096$ and higher $V_{0}^{*}$,
probably for the same reasons as those mentioned in the discussion
of the Debye-Waller correlation function.
In the variable box there was no such problem.
These features were the reason for us to use mainly the fixed box 
for $V_{0}^{*}<0.2$ and the variable box for
$V_{0}^{*}\ge0.2$. By the way, the same 'Debye-Waller problem' also occurs when
simulating hard disks in external periodic potentials, $N\ge4096$
and $V_{0}^{*}$ medium or high, and can be solved again by
using the variable box geometry.

However, for all system sizes the structure of the 
phase diagram with a pronounced
minimum at intermediate values of $V_0^\ast$ is not affected by the
shifts.

The difference in the value of $\kappa_{G}a_{s}$ at the transition
between the infinite and zero external potential cases, we find
$\kappa_{G}a_{s}(V_{0}^{*}=\infty)-\kappa_{G}a_{s}(V_{0}^{*}=0)
\approx0.82$. This is not far away from $0.608$ which is a value 
predicted by \cite{FNR}. 

We have also done simulations with slightly altered parameters, i.e using
particles with diameter $2R=3\mu m$, effective surface charge
$Z^{*}=20000$ and $a_{s}=8\mu m$, to match the experiments in \cite{BBL}.
As expected, we only observe a slightly shift of the phase diagram of
$\Delta(\kappa a_{s})\approx0.35$ towards higher values of $\kappa
a_{s}$. The experimental phase diagram \cite{BBL} qualitatively has
the same shape as our results, but shows larger freezing- and 
reentrance regions and is shifted to higher values of $\kappa
a_{s}$ at about $\Delta(\kappa a_{s})\approx4.5$ in average.
The reasons for these differences are probably due to the 
particle interaction. We only use pairwise interaction, which
is a good approximation for low particle densities \cite{BBSLG}. But for
higher particle densities many body interactions play a role because of
macroion screening, which results in an effective pair potential
that has considerable deviations \cite{BBSLG} from a pure Yukawa-like
potential like ours. In particular, there could be an attractive part.

\subsection{Scaling behavior}

We next try to determine the order of the phase transitions encountered in this
system for two values of $V_0^*$.  In order to investigate this 
issue we studied the scaling behavior
of the order parameter, susceptibility and the order parameter cumulant
near the phase boundary for a small ($2$) and a large ($1000$)
$V_0^{*}$.

From finite size scaling theory (for an overview
see Ref.~\cite{LB}) we expect these quantities to 
scale as~\cite{DL2} :
\begin{equation}
<\psi_{G_1}>_L L^b \sim f(L/\xi)
\label{scop}
\end{equation}
\begin{equation}
\chi_L k_BT L^{-c} \sim g(L/\xi)
\label{scchi}
\end{equation}
\begin{equation}
U_{L} \sim h(L/\xi)
\label{sccu}
\end{equation}
Here $b=\beta/\nu$, $c=\gamma/\nu$ (for critical scaling),
and $f$, $g$, $h$ are scaling functions.
Defining $\tilde{\kappa}=(\kappa a_{s}-\kappa_{G} 
a_{s})/\kappa_{G} a_{s}$, we
expect the correlation length $\xi$ to diverge as
$\xi \propto \tilde{\kappa}^{-\nu}$ for an ordinary critical point, while for
a KT-transition we have an essential singularity and
$\xi \propto \exp(a \tilde{\kappa}^{-\tilde{\nu}})$ when approaching
the transition from the liquid side.

In Fig.~\ref{sca} we have plotted the left 
hand sides of Eqs.(\ref{scop}), (\ref{scchi}) and (\ref{sccu}) versus $L/\xi$ for 
$V_0^*=1000$, where data points of the variable box geometry 
for $15.2 \leq \kappa a_{s} \leq 16.0$ have been 
considered and $\kappa_{G} a_{s} = 15.1$, obtained by cumulant intersection. In order 
not to introduce an unwarranted 
bias, we have separately considered ordinary critical scaling (left
column) and a KT scaling form (right column) and adjusted the values of the parameters 
$b$, $c$ and $\nu$, or $a$, $b$, $c$ and $\tilde{\nu}$, till we obtained
collapse of our data onto a single curve determined by a least square estimator.
\footnote{One remark concerning the KT-scaling: the errors in $\tilde{\nu}$ and
$a$ are relatively big because for example increasing $\tilde{\nu}$ and
decreasing $a$ by an appropriate amount resulted in a nearly equally
good collapse.}
Good collapse of our data is observed for 
both scaling forms, the numerical values for $\tilde{\nu}$, $2b=\eta$ and 
$c=2-\eta$ for KT scaling ($2b \approx 0.28$, $c \approx 1.70$, $\tilde{\nu} 
\approx 0.37$) are relatively close to the predicted values~\cite{FNR} 
($2b = \eta = 1/4$, $c = 1.75$, $\tilde{\nu} = 0.5$).
The situation is similar for small $V_0^* = 2$, the quality of the 
collapse comparable to the one of $V_{0}^{*}=1000$.
The critical parameters were obtained 
in this case for values  $15.4 \leq \kappa a_{s} \leq 16.2$,
with  $\kappa_{G} a_{s} = 15.37$.

We have a good
internal consistency between $\eta=0.25\dots0.33$ extracted from the
Debye-Waller correlation function, and the values obtained from data
collapsing: $\eta=0.28$ for $V_{0}^{*}=1000$, and $\eta=0.36$ for
$V_{0}^{*}=2$. 
Our results for the numerical values of the parameters are summarized in
Table~I.

A more precise classification of the phase transitions 
with the present data and system sizes is not easy.
This topic is left for future work, in particular 
we plan to compute the elastic properties of the system by a method
recently developed for the hard disk system~\cite{SNB,SNRB} and to test the 
KT predictions~\cite{FNR}.

\section{Summary and conclusion}
\label{concl}
In summary, we have calculated the phase diagram of a two dimensional system 
of soft disks, interacting via a DLVO potential, in an external
sinusoidal potential. We find  
freezing followed by reentrant melting transitions over a significant 
region of the phase diagram in tune with results on hard disks
\cite{SSN}, previous experiments on 
colloids~\cite{bech,BWL,BBL} and with the expectations of a dislocation
unbinding theory~\cite{FNR}. One of the main features of our calculation
is the method used to locate phase boundaries. In contrast to 
earlier simulations~\cite{CKSS,DK,DSK,DSK2} which used either the jump of 
the order parameter or
specific heat maxima to locate the phase transition, we used the 
more reliable cumulant intersection method. It must be noted that the 
specific heat in this system does not show a strong peak at the phase 
transition density so that its use 
may lead to confusing results. This, in our opinion, may be the reason for 
part of the controversy in this field. It is possible that earlier simulations
which used smaller systems and no systematic finite size analysis may have 
overlooked this feature of $c_V$ which becomes apparent only in computations 
involving large system sizes. We have shown that finite size scaling
of the order parameter cumulants as obtained from subsystem or
subblock analysis, on the other hand, yields an accurate phase 
diagram.

What is the order of the phase transition? We know that~\cite{henning,Jas,SNB} 
for the pure hard disk system in two dimensions this question is 
quite difficult to answer and our present understanding~\cite{SNB} is that 
this system shows a KTHNY transition.
In our system of soft disks, for zero external potential we can rule
out a strong first order transition, although smaller systems show
a feature (double peak in the internal energy) which mimics such a
behavior. We find several features which are consistent with the
KT theory, but also one which is not.
Upon turning on the external periodic potential, the difference
between hexatic and liquid disappears, and an (anisotropic) KT transition~\cite{FNR} 
from the modulated liquid into the solid is expected. 
Our results show several features which suggest that this is what we 
have, but there are still some (not large) deviations from theory.
Though we have discussed these observations in the rest of the paper, 
we list the important ones below for clarity:

\begin{itemize}

\item
The behavior of the cumulants near the transition 
is similar to an earlier work~\cite{DL} on the 
XY system which shows a KT transition. 

\item
The specific heat is relatively featureless and 
does not scale with system size in a fashion expected of a true first order or 
conventional continuous transition. 

\item
The decay of the correlation functions is similar to what is 
predicted~\cite{FNR} for an anisotropic scalar Coulomb gas.

\item
For two test values of $V_{0}^{*}$, the scaling of the order parameter, the 
susceptibility and the cumulant may be reasonable described by the KT
theory. 

\end{itemize} 

Of course, in order to resolve this issue unambiguously yet larger simulations
are required. Also, we need to compute elastic properties~\cite{SNRB,SNB} of
this system in order to compare directly with the results of Ref.~\cite{FNR}.
Work along these lines is in progress. 


\section{Acknowledgment}
We are grateful for many illuminating discussions with C. Bechinger 
and K. Binder.
One of us (S.S.) thanks the Alexander von Humboldt Foundation for
a Fellowship.
Support by the SFB~513 and granting of computer time
from the NIC and the HLRS is gratefully acknowledged.

\begin{figure}[hbtp]
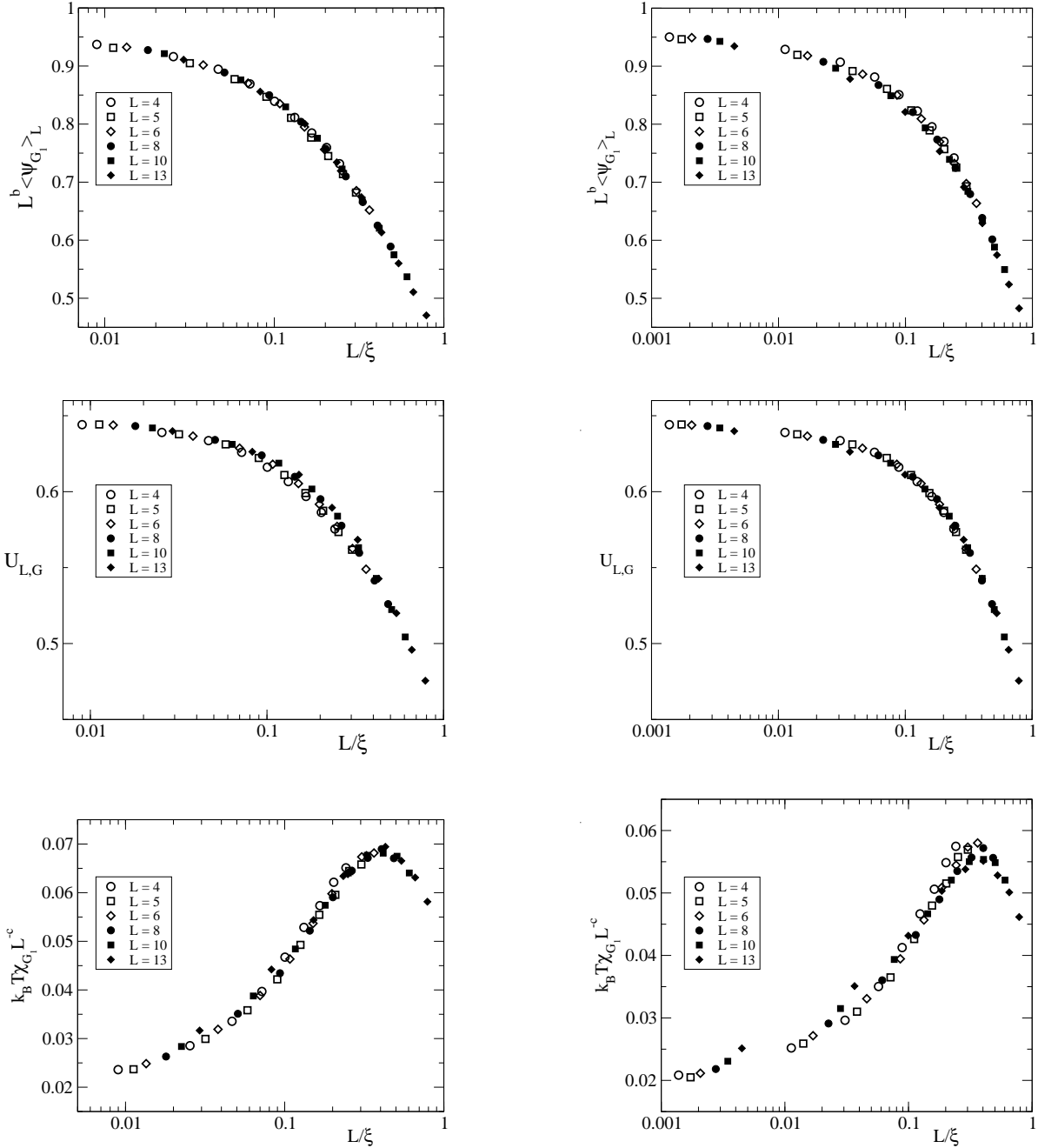

\begin{picture}(0,190)
\put(-10,105) {\psfig{figure=figures/N1024_V1000_DLVO2_mittel_kollaps_cr.eps,width=70mm}}
\put(-10,45) {\psfig{figure=figures/N1024_V1000_DLVO2_kumul_kollaps_cr.eps,width=70mm}}
\put(-10,-15) {\psfig{figure=figures/N1024_V1000_DLVO2_varianz_kollaps_cr.eps,width=70mm}}
\put(80,105) {\psfig{figure=figures/N1024_V1000_DLVO2_mittel_kollaps_KT.eps,width=70mm}}
\put(80,45) {\psfig{figure=figures/N1024_V1000_DLVO2_kumul_kollaps_KT.eps,width=70mm}}
\put(80,-15) {\psfig{figure=figures/N1024_V1000_DLVO2_varianz_kollaps_KT.eps,width=70mm}}
\end{picture}
\vskip 1.0 cm
\caption[]
{Scaling plots for the order parameter (first line),
the order parameter cumulant (second line) and the order parameter
susceptibility (third line) 
for $V_0^* = 1000$ assuming critical (left column) and KT scaling
(right column).
The total system size is $N = 1024$,
for $\xi$ we have used the expressions given after Eq.(\ref{sccu}).
}
\label{sca}
\end{figure}

\newpage
\begin{tabular} {||c||c||c|c|c||c|c|c|c||} \hline
 & $\kappa_{G} a_s$ & $b$ & $c$  & $\nu$ & $b$ & $c$  &  $\tilde{\nu}$ & $a$ \\ \hline
$V_0^*=1000$ & 15.1 & 0.130(12) & 1.61(8) & 1.5(2)& 0.14(1) & 1.70(3) & 0.37(6) & 1.45(40) \\
$V_0^*=2$ & 15.37 & 0.163(15) & 1.68(5) & 1.51(25)& 0.18(3) & 1.74(3) & 0.40(5) & 1.2(3) \\ \hline
KT theory& & & & & 0.125 & 1.75 & 0.5 & $O(1)$ \\ \hline
\end{tabular}
\vskip .5cm
\noindent
{\bf Table~I}~~ 
Parameters in the scaling plots (see Fig.~(\ref{sca}))
for $V_0^*=2$ and $V_0^*=1000$.
The first three parameter columns are for critical scaling, the last
four for KT scaling. The last line shows the predictions of KT theory.

\newpage

\noindent

\newpage

\newpage

\newpage

\newpage

\newpage

\newpage

\begin{figure}[hbtp]
\begin{picture}(0,80)
\put(-10,0) {\psfig{figure=figures/psiG_mittel_V1000_N4096_DLVO.eps,width=80mm}}
\put(80,0) {\psfig{figure=figures/psiG_kumul_V1000_N4096_DLVO2.eps,width=80mm}}
\end{picture}
\vskip .5 cm
\caption[]
{
Average (left) and cumulant (right) of the $\psi_{G_1}$ order parameter versus
$\kappa a_s$ for $V_0^* = 1000$ and various system sizes $L$ (N=4096).
}
\label{opkumv1000}
\end{figure}
\newpage

\newpage

\newpage

\newpage

\newpage

\begin{references}
\bibitem{CAK} N.A. Clark, B.J. Ackerson, A. J. Hurd,
\prl {\bf 50}, 1459 (1983).

\bibitem{CAC} A. Chowdhury, B.J. Ackerson, N. A. Clark,
\prl {\bf 55}, 833 (1985).

\bibitem{LA} K. Loudiyi, B. J. Ackerson, Physica {\bf A 184}, 1 (1992); 
{\it ibid} 26 (1992).

\bibitem{bech} Q.-H. Wei, C. Bechinger, D. Rudhardt and P. Leiderer, \prl
{\bf 81}, 2606 (1998).

\bibitem{BWL} C. Bechinger, Q.H. Wei, P. Leiderer, J. Phys.: Cond. Mat. 
{\bf 12}, A425 (2000).

\bibitem{BBL} C. Bechinger, M. Brunner, P. Leiderer, \prl {\bf 86}, 930 (2001)

\bibitem{maret} K. Zahn, R. Lenke and G. Maret, \prl {\bf 82}, 2721, (1999)

\bibitem{CKS} J. Chakrabarti, H.R. Krishnamurthy, A. K. Sood, \prl {\bf 73},
2923 (1994).

\bibitem{FNR} E. Frey, D. R. Nelson, L. Radzihovsky, \prl {\bf 83}, 2977 (1999).\\
L. Radzihovsky, E. Frey, D. R. Nelson, Phys. Rev. {\bf E63}, 031503 (2001)

\bibitem{CKSS} J. Chakrabarti, H.R. Krishnamurthy, A.K. Sood, S. Sengupta,
\prl {\bf 75}, 2232 (1995).

\bibitem{DK} C. Das, H.R. Krishnamurthy, \prb {\bf 58}, R5889 (1998).

\bibitem{DSK} C. Das, A.K. Sood, H.R. Krishnamurthy,
Physica {\bf A 270}, 237 (1999).

\bibitem{DSK2} C. Das, P. Chaudhuri, A. Sood, H. Krishnamurthy, 
Current Science, Vol. 80, No. 8, p. 959 (April 2001)

\bibitem{SSN}
W. Strepp, S. Sengupta, P. Nielaba, Phys. Rev. {\bf E63}, 046106 (2001).

\bibitem{BBSLG} M. Brunner, C. Bechinger, W. Strepp, V. Lobaskin,
  H.H. von Gruenberg, Europhys. Lett. 58 (6) , pp. 926-965 (2002)

\bibitem{col} For an introduction to phase transitions in colloids see,
A. K. Sood in {\it Solid State Physics}, E. Ehrenfest and D. Turnbull Eds.
(Academic Press, New York, 1991); {\bf 45}, 1; P. N. Pusey in {\it Liquids,
Freezing and the Glass Transition}, J. P. Hansen and J. Zinn-Justin Eds.
(North Holland, Amsterdam, 1991).

\bibitem{WB} K. W. Wojciechowski and A. C. Bra\'nka, Phys. Lett. {\bf 134A},
314 (1988).

\bibitem{henning} H. Weber, D. Marx and K. Binder, \prb {\bf 51}, 14636 (1995);
H. Weber and D. Marx, Europhys. Lett. {\bf 27}, 593 (1994).

\bibitem{Jas} A. Jaster, Phys. Rev. E {\bf 59}, 2594 (1999).

\bibitem{SNB} S. Sengupta, P. Nielaba, K. Binder,
\pre {\bf 61}, 6294 (2000).
 
\bibitem{metro} N. Metropolis, A. W. Rosenbluth, M. N. Rosenbluth, 
A. H. Teller, E. Teller, J. Chem. Phys. {\bf 21}, 1087 (1953).

\bibitem{LB} D.P. Landau, K. Binder, {\it A Guide to Monte Carlo
Simulations in Statistical Physics}, Cambridge University Press (2000).

\bibitem{aldwain} B. J. Alder and T. E. Wainwright, Phys. Rev. {\bf 127}, 359
(1962).

\bibitem{NH}
D. Nelson, B. Halperin, Phys. Rev. B {\bf 19}, 2457 (1979).

\bibitem{KTHNY} 
J. M. Kosterlitz, D. J. Thouless, J. Phys. {\bf C 6}, 1181 (1973)\\ 
B.I. Halperin and D.R. Nelson, \prl {\bf 41}, 121 (1978)\\
A.P. Young, Phys. Rev. B {\bf 19}, 1855 (1979)\\
K.J. Strandburg, Rev. Mod. Phys. {\bf 60}, 161 (1988)\\
H. Kleinert, {\em Gauge Fields in Condensed Matter}, Singapore,
World Scientific (1989)

\bibitem{ecstrand} K. J. Strandburg, \prb {\bf 34}, 3536 (1986).

\bibitem{SNRB} S. Sengupta, P. Nielaba, M. Rao and K. Binder,
\pre {\bf 61}, 1072 (2000).

\bibitem{KB} K. Binder, Z. Phys. {\bf B43}, 119 (1981);
K. Binder, \prl {\bf 47}, 693 (1981).

\bibitem{VRSB} K. Vollmayr, J.D. Reger, M. Scheucher, K. Binder,
Z. Phys. {\bf B 91}, 113 (1993).

\bibitem{DL} 
D.P. Landau, J. Magn. Magn. Mat. {\bf 31-34}, 1115 (1983))
 
\bibitem{DL2}
D.P. Landau, Phys. Rev. {\bf B27}, 5604 (1983).


\end{references}
\end{document}